# Unified Error Rates Analysis of MIMO Space-Time Block Codes over Generalized Shadowed $\kappa$-$\mu$ and $\eta$-$\mu$ Fading and AWGGN

Ehab Salahat, *Member, IEEE*, Ali Hakam, *Member, IEEE*
{ehab.salahat,ali.hakam.a}@ieee.org

*Abstract*—This paper presents a novel unified performance analysis of Space-Time Block Codes (STBCs) operating in the Multiple Input Multiple Output (MIMO) network and subjected to generalized shadowed fading and noise scenarios. Specifically, we derive novel, simple and accurate average bit error rates (ABER) expressions for coherent modulation schemes in the generalized $\eta - \mu$ and shadowed $\kappa - \mu$ fading channels. The noise in the network is assumed to be modeled by the additive white generalized Gaussian noise (AWGGN), which encompasses the Laplacian and the Gaussian noise environments as special cases. The result obviates the need to re-derive the error rates for MIMO STBC systems under many multipath fading and noise models, while avoiding computationally-expensive expressions. Published results from the literature as well as numerical simulations corroborate the accuracy of our derived generalized expressions.

## I. INTRODUCTION

NEXT GENERATION mobile communication is seen to increase data transfer rate in bandwidth-limited wireless channels. Wireless channels, however, suffer from time-varying impairments that degrade the transmission quality and the performance, e.g. multipath fading and noise [1] [2] [3]. Space-time coding is a famous technique utilized to solve bandwidth limitation and its associated problems based on Orthogonal Frequency Division Multiplexing (OFDM) [4]. Compared with the other coded communications system, space-time coding can effectively improve the capacity of wireless transmission in relatively simple receiver structure, and get a higher coding gain without sacrificing the whole bandwidth [5]. Moreover, using multiple transmitter and/or receiver antennas in MIMO systems enhances the reliability of wireless transmission significantly due to property of diversity [6]. Maximal ratio, equal gain, and selection combining techniques are well-known receiver diversity which are used in different mobile communication systems. On the other hand, transmitter diversity is more advantageous as the gain due to diversity is obtained for downlink transmission without using multiple antennas at the receiver. Space-time block codes (STBCs), introduced by Alamouti [7], Tarokh [8] and many other wireless pioneer researchers, are used in most MIMO systems, as they are considered the standard approach to combat channel fading due to their transmit code-words simplicity, receiver's low complexity, the ability to achieve full-diversity and enhance the end-to-end wireless network performance [9] [10].

The literature is rich with research work on STBCs in MIMO fading channels that have led to better understanding of their expected performance and design trade-offs. For example, an excellent survey paper on signal identification on MIMO and STBC can be found in [11]. The authors in [12] proposed a new orthogonal STBC scheme that can resolve the linear decoding problem, keep the simplicity and high performance properties of the classical OSTBC, and achieve the query diversity for the MIMO backscatter RFID.

The MIMO channels, however, are generally assumed to be independent and identically distributed (i.i.d.) Rayleigh or Rician fading channels. For example, the work in [13] presents error analysis of multiple rate STBC for MIMO network over Rician fading channel. In [14], the authors assume Nakagami-$m$ fading for their MIMO system for simplicity. Although there is no doubt such assumption comes with a certain mathematical convenience, it may not be physically justified for many scenarios of interest, or at least these assumptions limit their wider applicability. Most of the previous research work was evaluated for the case of short-term multipath fading only, while ignoring the large-scale fading effect (shadowing). To circumvent on this issue, some published worked considered to use the generalized-$K$ model. For instance, the study of the error exponent of generalized-$K$ MIMO channels over orthogonal STBC was considered in [15] and [16]. The channel capacity and the error rates expressions were also derived in [17] and [18], where the results were obtained in terms of complicated special functions. A simpler shadowed-fading model, the Rayleigh-Lognormal, was also considered to numerically evaluate the outage capacity.

Such inefficient expressions, even though being featured with a very limited generality, are not easily amenable for further manipulations. Above all the discussed issues, all published work in the literature treat the calculations assuming Gaussian noise environment, i.e. the additive white Gaussian noise (AWGN) model.

To circumvent on such issues, in this paper we present a novel unified performance analysis of Space-Time Block Codes (STBC) operating in the MIMO network. Novel, simple and accurate ABER expressions are derived in the generalized $\eta$-$\mu$ and shadowed $\kappa$-$\mu$ fading channels. The noise in the network is assumed to be additive white generalized Gaussian noise (AWGGN), which encompasses many noise special cases e.g. the Laplacian noise and the AWGN as special cases. The new expressions allow the performance of many wireless systems to be evaluated under many fading and noise scenarios. The result obviates the need to re-derive similar results for the special cases, encompassed by our new analytical expressions in a piecemeal fashion. Published results from the literature as well as numerical solutions corroborate the accuracy of our new generalized ABER expression.

The remaining sections of this work are structured as follows. Section II introduces the mathematical derivations and the associated system model, fading channels, and noise environment. Derivation of the unified error rates expressions are presented in section III with their simulation results in section IV. Finally, the paper findings and conclusion are briefed in section V.

## II. MATHEMATICAL DERIVATIONS

### A. The $\eta - \mu$ and $\lambda - \mu$ Fading Models

The power Probability Density Function (PDF) of the $\eta - \mu$ and $\lambda - \mu$ (or $\eta - \mu$ format II) generalized fading models is given in [19] [20] by:

$$f_\gamma(\gamma) = \frac{2\sqrt{\pi}\mu^{\mu+0.5}h^\mu}{\Gamma(\mu)H^{\mu-0.5}\tilde{\gamma}^{\mu+0.5}}\gamma^{\mu-0.5}e^{-\left[\frac{2\mu h}{\tilde{\gamma}}\right]\gamma}I_{\mu-0.5}\left(\left[\frac{2\mu H}{\tilde{\gamma}}\right]\gamma\right), \quad (1)$$

where $I_v(\cdot)$ is the modified Bessel function of the first type [21] and order $v$, and $h$ and $H$ are shown in Table I for the two fading models. In these two generalized fading models, the physical parameters $\lambda$, $\eta$, and $\mu$ account for the correlation between the in-phase and quadrature components of the fading signal (assuming each is a zero-mean), the unequal power of these components, and the number of multipath clusters [19].

TABLE I: $h$ AND $H$ VALUES FOR THE $\eta - \mu$ AND $\lambda - \mu$ MODELS [2]

| Distribution | $h$ | $H$ |
| --- | --- | --- |
| $\eta - \mu$ | $0.25\eta^{-1}(1+\eta)^2$ | $0.25\eta^{-1}(1-\eta^2)$ |
| $\lambda - \mu$ | $(1-\lambda^2)^{-1}$ | $\lambda(1-\lambda^2)^{-1}$ |

This generalized fading distribution includes the Nakagami-$m$, the Rice, the Hoyt, the Rayleigh and the One-Sided Gaussian models as special cases. Table I in [2] and table II in [22] summarizes how the special cases of these generalized fading models can be obtained from (1).

### B. The $\kappa - \mu$ Shadowed Fading Model

The $\kappa-\mu$ shadowed fading model relies on a generalization of the physical model of the $\kappa-\mu$ distribution [19]. Assuming a non-homogeneous propagation environment, the multipath waves are assumed to have scattered waves with the same power and an arbitrary power for the dominant component. In contrast with the $\kappa-\mu$ distribution which assumes a deterministic dominant component within each cluster, the $\kappa-\mu$ shadowed model assumes that the dominant components of the clusters can change randomly due to shadowing. Since the $\kappa-\mu$ distribution includes the Rician distribution as a special case [19] [23], a natural generalization of the $\kappa-\mu$ fading distribution can be obtained by line-of-sight (LOS) shadow fading model with the same multipath/shadowing scheme used in the Rician shadowed model [24]. Assuming that shadowing follows Nakagami-$m$, then the probability density function of the $\kappa-\mu$ shadowed fading is given as [24]:

$$f_\gamma(\gamma) = \frac{\mu^\mu m^m (1+\kappa)^\mu}{\Gamma(\mu)(\mu\kappa+m)^m \tilde{\gamma}^\mu}\gamma^{\mu-1}e^{-\left[\frac{\mu(1+\kappa)}{\tilde{\gamma}}\right]\gamma}{}_1F_1\left(m;\mu;\left[\frac{\mu^2\kappa(1+\kappa)}{[\mu\kappa+m]\tilde{\gamma}}\right]\gamma\right), \quad (2)$$

where $\kappa$, $\mu$ and $m$ account for the ratio between the total power of the dominant components and the total power of the scattered waves, the number of the multipath clusters, and the Nakagami-$m$ shadowing parameter, respectively [24] [19], and the functions ${}_1F_1(\cdot)$ and $\Phi_2(\cdot)$, are as defined in [25]. The model includes the $\kappa-\mu$, the Nakagami-$m$, the Rician shadowed, the Rician, the Rayleigh and the one-sided Gaussian as special cases. These distributions can be obtained from (2) as shown in and summarized in Table II for brevity.

TABLE II: SPECIAL CASES OF THE $\kappa - \mu$ SHADOWED FADING [23].

| Fading Distribution | $\mu$ | $\kappa$ | $m$ |
| --- | --- | --- | --- |
| $\kappa - \mu$ | $\mu$ | $\kappa$ | $m \to \infty$ |
| Nakagami-$m$ | $\mu = m$ | $\kappa \to 0$ | $m \to \infty$ |
| Rician shadowed | $\mu = 1$ | $\kappa = K$ | $m = m$ |
| Rician | $\mu = 1$ | $\kappa = K$ | $m \to \infty$ |
| Rayleigh | $\mu = 1$ | $\kappa \to 0$ | $m \to \infty$ |
| One-Sided Gaussian | $\mu = 0.5$ | $\kappa \to 0$ | $m \to \infty$ |

### C. Space-Time Block Coding Analysis

Following the same analysis and the assumptions given in [26], and assuming that the number of transmit antennas and receive antennas is given respectively by $N_t$ and $N_r$, it was shown in [26] that the received power PDF (uncorrelated antennas) for the $\eta - \mu$ generalized fading is given by [22]:

$$f_\gamma(\gamma) = \frac{2\sqrt{\pi}h^{\mu N_t N_r}}{\Gamma(\mu N_t N_r)}\left(\frac{\mu}{\tilde{\gamma}}\right)^{\mu N_t N_r + 0.5}\left(\frac{\gamma}{H}\right)^{\mu N_t N_r - 0.5} \times e^{-\left[\frac{2\mu h}{\tilde{\gamma}}\right]\gamma}I_{\mu N_t N_r - 0.5}\left(\left[\frac{2\mu H}{\tilde{\gamma}}\right]\gamma\right), \quad (3)$$

which, for compactness, is written as:

$$f_\gamma(\gamma) = \psi\gamma^{m-1}e^{-\beta\gamma}I_v(\xi\gamma), \quad (4)$$

where $\psi = \frac{2\sqrt{\pi}h^{\mu N_t N_r}}{\Gamma(\mu N_t N_r)H^{m-1}}\left(\frac{\mu}{\tilde{\gamma}}\right)^m$, $m = \mu N_t N_r + 0.5$, $\beta = \left[\frac{2\mu h}{\tilde{\gamma}}\right]$, $\xi = \left[\frac{2\mu H}{\tilde{\gamma}}\right]$ and $v = m - 1$.

Similarly, it was shown in [24] that the sum of $L$ i.i.d. generalized $\kappa - \mu$ shadowed random variables (R.V.) with parameters $\kappa$, $\mu$, $m$ and $\tilde{\gamma}$ is also another $\kappa - \mu$ shadowed R.V. with the parameters $\kappa$, $L\mu$, $Lm$ and $L\tilde{\gamma}$. Following this result, the received power PDF with $N_t$ transmit antenna and $N_r$ received antenna can be expressed as:

$$f_\gamma(\gamma) = \frac{\tilde{\mu}^{\tilde{\mu}}\tilde{m}^{\tilde{m}}(1+\kappa)^{\tilde{\mu}}}{\Gamma(\tilde{\mu})(\tilde{\mu}\kappa+\tilde{m})^{\tilde{m}}\eta^{\tilde{\mu}}}\gamma^{\tilde{\mu}-1}e^{-\left[\frac{\tilde{\mu}(1+\kappa)}{\eta}\right]\gamma}{}_1F_1\left(\tilde{m};\tilde{\mu};\left[\frac{\tilde{\mu}^2\kappa(1+\kappa)}{[\tilde{\mu}\kappa+\tilde{m}]\eta}\right]\gamma\right), \quad (5)$$

where $\tilde{\mu} = N_t N_r \mu$, $\tilde{m} = N_t N_r m$, $\eta = N_t N_r \tilde{\gamma}$, which is written (for compactness) in the form:

$$f_\gamma(\gamma) = \psi\gamma^{\tilde{\mu}-1}e^{-\beta\gamma}{}_1F_1(\tilde{m};\tilde{\mu};\zeta\gamma), \quad (6)$$

where $\psi = \frac{\tilde{\mu}^{\tilde{\mu}}\tilde{m}^{\tilde{m}}(1+\kappa)^{\tilde{\mu}}}{\Gamma(\tilde{\mu})(\tilde{\mu}\kappa+\tilde{m})^{\tilde{m}}\eta^{\tilde{\mu}}}$, $\beta = \left[\frac{\tilde{\mu}(1+\kappa)}{\eta}\right]$, and $\zeta = \left[\frac{\tilde{\mu}^2\kappa(1+\kappa)}{[\tilde{\mu}\kappa+\tilde{m}]\eta}\right]$.

The expressions in (4) and (6) will be used later in our unified bit error rates analysis.

## D. The AWGGN Environment

For simplicity, many researchers consider the AWGN model as their noise conditions, given in terms of the famous Gaussian $Q$-function. Wireless communications channels might be subjected to non-Gaussian noise. Examples of the non-Gaussian noise include the Laplacian and the Gamma noise (see [1] [3] [27] [28] for more discussion on the different noise environments). The additive white generalized Gaussian noise (AWGGN) is a generalized noise model that renders the aforementioned noise models as special cases. It is written using the generalized $Q$–function, and is given by:

$$Q_a(x) = \frac{a\Lambda_0^{2/a}}{2\Gamma(1/a)} \int_x^\infty e^{-\Lambda_0^a |u|^a} du = \frac{\Lambda_0^{2/a-1}}{2\Gamma(1/a)} \Gamma(1/a, \Lambda_0^a |x|^a). \quad (7)$$

where $\Lambda_0 = \sqrt{\Gamma(3/a)/\Gamma(1/a)}$, $\Gamma(\cdot)$ denotes the famous gamma function, and $a$ being the noise parameter. Table III illustrates how the noise special cases can be achieved from (7).

TABLE III: Relation Between $Q_a(x)$ and its Special Cases.

| Noise Dist. | Impulsive | Gamma | Laplacian | Gaussian | Uniform |
|---|---|---|---|---|---|
| $a$ | 0.0 | 0.5 | 1.0 | 2.0 | $\infty$ |

It was shown in [1] that (7) can be efficiently and accurately approximated (the robustness of the approximation is shown in [1]) as a simple summation of four decaying exponential functions, given as:

$$Q_a(\sqrt{x}) \approx \sum_{i=1}^4 p_i e^{-q_i x}, \quad (8)$$

where the fitting parameters, $p_i$ and $q_i$ are obtained using nonlinear curve fitting (Levenberg-Marquardt method), with sample fitting values for different cases of the noise parameter $a$ being presented in Table IV [1].

TABLE IV: Fitting Parameters of $Q_a(\sqrt{\cdot})$ Approximation

| $a$ | $p_1$ | $p_2$ | $p_3$ | $p_4$ | $q_1$ | $q_2$ | $q_3$ | $q_4$ |
|---|---|---|---|---|---|---|---|---|
| 0.5 | 44.920 | 126.460 | 389.400 | 96.540 | 0.130 | 2.311 | 12.52 | 0.629 |
| 1 | 0.068 | 0.202 | 0.182 | 0.255 | 0.217 | 2.185 | 0.657 | 12.640 |
| 1.5 | 0.065 | 0.149 | 0.136 | 0.125 | 0.341 | 0.712 | 10.57 | 1.945 |
| 2 | 0.099 | 0.157 | 0.124 | 0.119 | 1.981 | 0.534 | 0.852 | 10.268 |
| 2.5 | 0.126 | 1.104 | -1.125 | 0.442 | 9.395 | 0.833 | 0.994 | 1.292 |

This approximation will be used in section III in our unified performance expression.

## III. The Unified Bit Error Rates Analysis

### A. Average Bit Error Rate

The average bit error rate due to a fading channel can be evaluated by averaging the bit error rate of the noisy channel using the fading PDF [29] [30]. This averaging process with $f_\gamma(\gamma)$ representing the fading PDF and AWGGN environment can be expressed as:

$$P_e = \mathcal{A} \int_0^\infty f_\gamma(\gamma) Q_a(\sqrt{\mathcal{B}\gamma}) d\gamma, \quad (9)$$

where $Q_a(\sqrt{\cdot})$ is the AWGGN (conditional error probability given the channel state), and $\mathcal{A}$ and $\mathcal{B}$ are modulation scheme dependent, as shown in Table V.

TABLE V: $\mathcal{A}$ and $\mathcal{B}$ Values For Different Modulations

| Modulation Scheme | Average SER | $\mathcal{A}$ | $\mathcal{B}$ |
|---|---|---|---|
| BFSK | $= Q_a(\sqrt{\gamma})$ | 1 | 1 |
| BPSK | $= Q_a(\sqrt{2\gamma})$ | 1 | 2 |
| QPSK, 4-QAM | $\approx 2Q_a(\sqrt{\gamma})$ | 2 | 1 |
| M-PAM | $\approx \frac{2(M-1)}{M} Q_a\left(\sqrt{\frac{6}{M^2-1}\gamma}\right)$ | $\frac{2(M-1)}{M}$ | $\frac{6}{M^2-1}$ |
| M-PSK | $\approx 2Q_a\left(\sqrt{2\sin^2\left(\frac{\pi}{M}\right)\gamma}\right)$ | 2 | $2\sin^2\left(\frac{\pi}{M}\right)$ |
| Rectangular M-QAM | $\approx \frac{4(\sqrt{M}-1)}{\sqrt{M}} Q_a\left(\sqrt{\frac{3}{M-1}\gamma}\right)$ | $\frac{4(\sqrt{M}-1)}{\sqrt{M}}$ | $\frac{3}{M-1}$ |
| Non-Rectangular M-QAM | $\approx 4Q_a\left(\sqrt{\frac{3}{M-1}\gamma}\right)$ | 4 | $\frac{3}{M-1}$ |

Using (9), we will derive the ABER expressions for the $\eta - \mu$ and $\kappa - \mu$ shadowed generalized fading models subjected to AWGGN next.

### B. $\eta - \mu$ fading

Substituting (4) into (9), then (9) can be re-written as:

$$P_e = \mathcal{A}\psi \int_0^\infty \gamma^{m-1} e^{-\beta\gamma} I_\nu(\xi\gamma) Q_a(\sqrt{\mathcal{B}\gamma}) d\gamma, \quad (10)$$

utilizing the generalized $Q_a(\sqrt{\cdot})$ approximation in (8), then (10) can then be written in the form:

$$P_e = \sum_{i=1}^4 [\mathcal{A}\psi p_i] \int_0^\infty \gamma^{m-1} e^{-\tilde{\beta}\gamma} I_\nu(\xi\gamma) d\gamma, \quad (11)$$

which can be evaluated in a closed-form [21], evaluating to

$$P_e = \sum_{i=1}^4 \Psi \,_2F_1\left(\left[\frac{m+\nu}{2}, \frac{m+\nu+1}{2}\right]; [1+\nu], \frac{\xi^2}{\tilde{\beta}^2}\right), \quad (12)$$

where $\tilde{\beta} = [\beta + q_i \mathcal{B}]$, $\Psi = \frac{[\mathcal{A}\psi p_i]\xi^\nu \Gamma(m+\nu)}{2^\nu \tilde{\beta}^{m+\nu}\Gamma(\nu+1)}$, and $_2F_1([\cdot,\cdot];\cdot;\cdot)$ is the Gauss hypergeometric function [21].

This new derived expression in (12) is simple, generic and generalized for the ABER analysis in the $\eta - \mu$ two formats deploying STBC reception in AWGGN, and applies directly to all the special cases of the $\eta - \mu$ fading model. This expression will be utilized in the illustration of the simulation in the following section.

### C. $\kappa - \mu$ shadowed fading

Substituting (6) into (9), then (9) can be rewritten as

$$P_e = \mathcal{A}\psi \int_0^\infty \gamma^{\tilde{\mu}-1} e^{-\beta\gamma} \,_1F_1(\tilde{m}; \tilde{\mu}; \zeta\gamma) Q_a(\sqrt{\mathcal{B}\gamma}) d\gamma, \quad (13)$$

and using the generalized $Q_a(\sqrt{\cdot})$ approximation from (8), one arrives at the integral

$$P_e = \mathcal{A}\psi \sum_{i=1}^4 p_i \int_0^\infty \gamma^{\tilde{\mu}-1} e^{-[\beta+q_i\mathcal{B}]\gamma} \,_1F_1(\tilde{m}; \tilde{\mu}; \zeta\gamma) d\gamma, \quad (14)$$

with simple variable transform, as $z = \zeta\gamma$, (15) is evaluated in a simple closed form as

$$P_e = \sum_{i=1}^4 \Psi \,_2F_1\left([\tilde{m}, \tilde{\mu}]; [\tilde{\mu}], \tilde{\beta}^{-1}\right), \quad (15)$$

with $\tilde{\beta} = [\beta + q_i\mathcal{B}]/\zeta$ and $\Psi = [\mathcal{A}\psi p_i \Gamma(\tilde{\mu})]/[\zeta^{\tilde{\mu}} \tilde{\beta}^{\tilde{\mu}}]$.

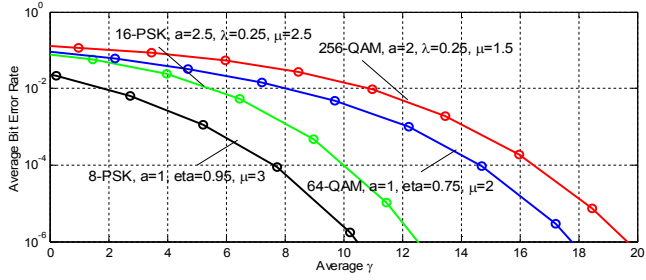

Fig. 1: ABER for different modulation schemes in generalized $\eta - \mu$ fading subjected to different types of noise ($N_t = N_r = 2$).

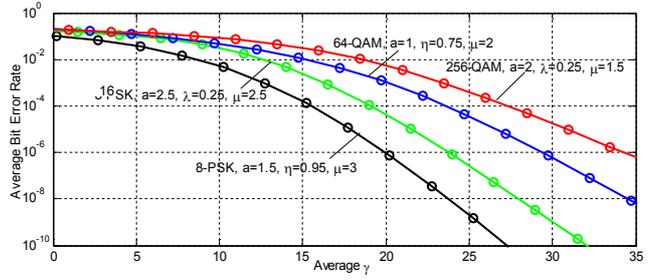

Fig. 2: ABER for different modulation schemes in generalized $\eta - \mu$ fading subjected to different types of noise ($N_t = 1, N_r = 1$).

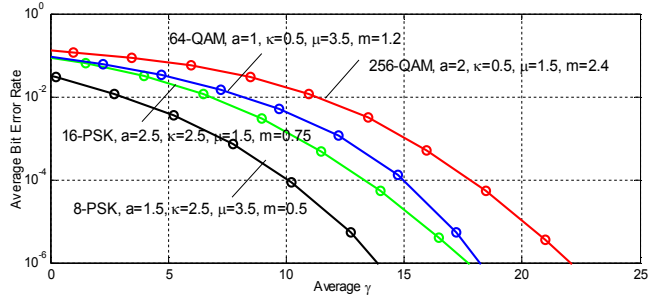

Fig. 3: ABER for different modulation schemes in generalized $\kappa - \mu$ shadowed fading subjected to different types of noise ($N_t = N_r = 2$).

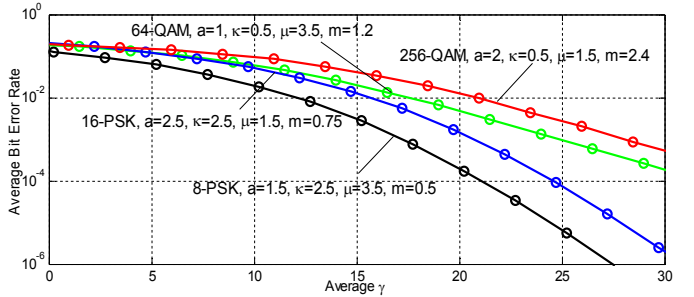

Fig. 4: ABER for different modulation schemes in generalized $\kappa - \mu$ shadowed fading subjected to different types of noise ($N_t = N_r = 1$).

This expression is new, and it generalizes the results to all the special cases of the $\kappa - \mu$ fading and AWGGN special cases, deploying STBC.

## IV. NUMERICAL RESULTS

This section illustrates sample ABER simulation results for the derived expressions in (12) and (15), which are compared with the numerically obtained results. Five test cases are presented, assuming different fading and noise conditions, with different modulation schemes, and testing codes are available in appendix A for the reader's reference. For the clarity of the plots in the figures, the use of legends is avoided, however, please note that the solid curves represent the numerical results whereas the overlaid patterns represent the derived expressions. Moreover, please note that the fading and noise conditions as well as the utilized modulation schemes are also indicated on the plots.

In the first and second tests, we consider the $\eta - \mu$ fading with $N_t = N_r = 2$ and $N_t = N_r = 1$, respectively. The generated results are shown in Fig. 1 and 2, respectively. One can clearly see the excellent match between the numerical results and those obtained using (12).

Similarly, the third and fourth test scenarios follow the same assumption of the first two tests, but assuming $\kappa - \mu$ shadowed fading with the parameters shown in the plots. The generated results, shown in Fig. 3 and 4 indicate a very good match between (15) and the numerically obtained results.

As expected, the performance gain due to the utilization of the multiple transmit and receive antenna can be clearly observed by comparing Fig. 1 vs. Fig. 2 and Fig. 3 vs. Fig. 4. Moreover, one can see the effect of the fading and noise parameters, by for example, comparing the enhanced performance with higher values of $\mu$ and the degraded performance due to the low values of $a$ (see table III). Finally, as a further verification step, we regenerate in Fig. 5 the plots given in [20], Fig. 2, using (15), where one can see identical results.

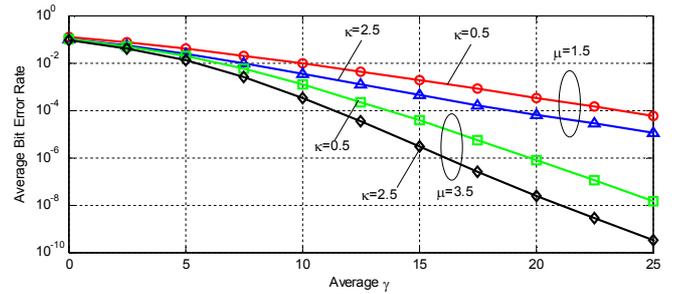

Fig. 5: Regenerated figure from ( [20] Fig.2).

## V. CONCLUSION

This paper presented novel unified performance analysis of STBC in MIMO network. The new derived expressions for the ABER are unified, simple and applicable for all coherent modulation schemes assuming $\eta - \mu$ and shadowed $\kappa - \mu$ fading channels and the AWGGN noise environment. The result obviates the need to re-evaluate similar expressions for many fading and noise models in a piecemeal fashion. Intensive testing proves the accuracy of our expressions.